\documentclass[11pt,preprint]{aastex}

\def\lya{Ly$\alpha$}
\def\lyaesc{$f^{Esc}_{Ly\alpha}$}

\def\ergcm2s{\ifmmode {\rm\,erg\,cm^{-2}\,s^{-1}}\else
                ${\rm\,ergs\,cm^{-2}\,s^{-1}}$\fi}
\def\ergsec{\ifmmode {\rm\,erg\,s^{-1}}\else
                ${\rm\,ergs\,s^{-1}}$\fi}
\newcommand{\fluxunit}{\ergcm2s}

\begin{document}

\title{X-ray Constraints on the \lya\ Escape Fraction}

\author{Zhen-Ya Zheng$^{1,2}$; Sangeeta Malhotra$^{2}$; Jun-Xian Wang$^{1}$; James E. Rhoads$^{2}$; 
Steven L. Finkelstein$^{3}$; Eric Gawiser$^{4}$; Caryl Gronwall$^{5}$; Lucia Guaita$^{6}$; Kim K. Nilsson$^{7}$; Robin Ciardullo$^{5}$}


\begin{abstract}
We have coadded the X-ray flux of all known \lya\ emitters in the 4 Msec
Chandra Deep Field South (CDF-S) region, achieving the tightest upper limits yet on
the X-ray to \lya\ ratio.  We use the X-ray data 
to place sensitive upper limits on the average unobscured star-formation 
rate (SFR$_X$)  in these galaxies. A very small fraction of \lya\ galaxies in the field are individually detected 
in the X-rays, implying a low fraction of AGN activity.  After excluding the  few X-ray detected Lyman alpha 
emitters (LAEs), we stack the undetected LAEs located in the 4 Ms CDF-S data and 250 ks Extended 
CDF-S (ECDFS) data, and  compute a 1-$\sigma$ upper limit on SFR$_X$ $<$ 1.6, 14, 28, 28, 140, 440, 
880 M$_{\odot}$ yr$^{-1}$ for LAEs located at z $\sim$ 0.3 and z = 2.1, 3.1, 3.2, 4.5, 5.7 and 6.5, respectively. The upper 
limit of SFR$_X$ in LAEs can be then be compared to SFR$_{Ly\alpha}$ derived from \lya\ line and thus can constrain 
on the \lya\ escape fraction ($f^{Esc}_{Ly\alpha}$). The \lyaesc\ from X-ray at z $\sim$ 0.3 is 
substantially larger than that from UV or H$\alpha$. Three X-ray detected LAE galaxies at z$\sim$ 0.3 show \lyaesc\ $\sim$ 3-22\%, 
and the average \lya\ escape fraction from stacking the X-ray undetected LAEs show \lyaesc\ $>$ 28\% at 3 $\sigma$ 
significance level at the same redshift. We derive a  lower limit on \lyaesc\ $>$ 14\% (84 \% confidence 
level, 1 $\sigma$ lower limit) for LAEs at redshift z $\sim $2.1 and  z $\sim$ 3.1-3.2. At z $>$ 4, the current
LAE samples are not of sufficient size to constrain SFR$_X$ well. By averaging all the LAEs at z$ >$ 2, 
the X-ray non-detection constrains \lyaesc\ $ >$ 17\% (84 \% confidence level, 1 $\sigma$ lower limit), 
and rejects \lyaesc\ $<$ 5.7\% at the 99.87\% confidence level from 2.1 $<$ $z$ $<$ 6.5. 
\end{abstract}


\altaffiltext{1}{Department of Astronomy, University of Science and Technology of China, Hefei, Anhui 230026, China; zhengzy@mail.ustc.edu.cn} 

\altaffiltext{2}{School of Earth and Space Exploration, Arizona State University, Tempe, AZ 85287}

\altaffiltext{3}{George P. and Cynthia Woods Mitchell Institute for Fundamental Physics and Astronomy, Department of Physics and Astronomy, Texas A\&M University, College Station, TX 77843} 

\altaffiltext{4}{Department of Physics and Astronomy, Rutgers University, 136 Frelinghuysen Rd., Piscataway, NJ 08854}

\altaffiltext{5}{Department of Astronomy \& Astrophysics, Penn State University, State College, PA 16802}

 \altaffiltext{6}{Department of Astronomy, Oscar Klein Center, Stockholm University, AlbaNova, Stockholm SE-10691, Sweden}

 \altaffiltext{7}{European Southern Observatory, Karl-Schwarzschild-StraBe 2, 85748 Garching bei Munchen, Germany}

\section {INTRODUCTION}

\lya\ emission, the 1216 \AA\ (n = 2$\rightarrow$1) transition of hydrogen in emission,  is a prominent 
tracer of ionizing photons produced by young stars. This line can carry up to $\sim$6\% of the 
bolometric luminosity of a star forming galaxy (Partridge \& Peebles 1967), therefore it is an easy 
handle for detection of both star forming galaxies and active galactic nuclei (AGN, where \lya\ 
emission is powered by the ionizing photons produced at the accretion disk around the central massive black hole) at redshifts 
$z > $ 2. The \lya\ line-search technique has been used successfully to identify samples of high redshift 
galaxies for over a decade, using narrowband images (e.g., Cowie \& Hu 1998; Rhoads et al. 2000, 
2003; Malhotra \& Rhoads 2002, Gawiser et al. 2007; Finkelstein et al. 2008, 2009; Guaita et al. 2010) and spectroscopic 
surveys (e.g., Pirzkal et al. 2004, Deharveng et al. 2008, Martin et al. 2008, Rhoads et al. 2009, and Blanc et al. 2010). 
There are thousands of photometrically selected \lya\ emitters (hereafter LAEs), with hundreds of spectroscopic confirmations (e.g., Hu et al. 2004, 
Dawson et al. 2007, Wang et al. 2009) at redshifts ranging from $z \approx$ 0.3 (Deharveng et al. 2008) to $z \approx$ 7 (Iye et al. 2006).

Since \lya\ line searches have achieved notable success in identifying high 
redshift star-forming galaxies, it is very important to understand the radiation transfer
of the \lya\ emission line. However, interpreting the \lya\ line is not trivial, because \lya\ photons are resonantly scattered 
when they interact with the surrounding neutral hydrogen in the inter-stellar medium (ISM). These 
radiative transfer effects can be quite complex when considering the presence of dust  in a multiphase 
and moving interstellar medium (Neufeld 1991, Hansen \& Oh 2006; Finkelstein et al. 2009c), 
and require empirical study of the fraction of \lya\ photons of which can escape the ISM.  
{\it Indeed, empirical results on the  \lya\ escape  fraction offer one of the better tools for probing the 
physics of \lya\ escape, which is determined on spatial scales that are usually unresolved in 
distant or nearby galaxies.}   On the theory
side, semi-analytical models of LAE populations by Le Delliou et al. (2005) assume a constant 
escape fraction of f$_{esc}$ = 0.02, and hydrodynamical models of LAEs at z $\sim$ 3 predict  f$_{esc}$ = 0.05-0.1 (Nagamine et al. 2010, Shimizu et al. 2011).

On the observational side, to measure the fraction of \lya\ photons
escaping the ISM (hereafter the \lya\ escape fraction), we often require a
flux measurement in ultraviolet (UV) continuum or a
non-resonant recombination line, such as H$\alpha$, and a dust extinction value. The dust corrected
UV luminosity or H$\alpha$ luminosity, like the intrinsic \lya\ luminosity, are mainly contributed by young stars, thus 
connected to the intrinsic star formation rate (SFR). Recent constraints at z $\sim$ 0.3 (Atek et al. 2009) and at z $\sim$2.25 (Hayes et al. 2010) are
based on the line ratio between H$\alpha$ and \lya. With the optical spectroscopic observation of GALEX selected LAEs, Atek et al. (2009) showed
that the \lya\ escape fractions span a wide range (from f$_{esc}$ $\sim$ 0.5 to 100 \%) and decrease with increasing dust extinction. 
Hayes et al. (2010) estimated an average \lya\ escape fraction of $\sim$ 5\% with a double blind survey targeting \lya\ and H$\alpha$ at z = 2.25. 
With the dust-corrected UV emission of 98 LAEs at z $\approx$ 1.9-3.8 detected through integral-field 
spectroscopy survey, Blanc et al. (2010) got
a median \lya\ escape fraction of 29\%. However, the escape fractions estimated from both of these methods are sensitive to
the dust extinction law. For example, under the case B recombination, if we ignore the 
resonant scattering and velocity field in the gas or ISM geometry, the \lya\ escape fraction is presented as
f$_{esc}$ = 10$^{-0.4\mbox{ }k(Ly\alpha)\mbox{ }E(B-V)}$ (Atek et al. 2009), and the extinction coefficient at \lya\ wavelength goes from 
$k(1216)$ $\sim$ 9.9 to 12.8 for Cardelli et al. (1989) and Calzetti et al. (2000) laws, respectively. This
will lead to a factor of two difference in the inferred f$_{esc}$ when E(B-V) $\geq$ 0.3. Meanwhile, 
the escape fractions estimated here also depend on the selection methods of LAEs. At z $\sim$ 0.3, the LAEs are selected 
from a spectroscopic followup of FUV dropouts (Atek et al. 2009), so the escape fractions here refer to a subsample of dropout galaxies at that redshift. At z $\sim$ 2.25, the average \lya\ escape fraction is estimated from 18 LAEs and 55 H$\alpha$ emitters, and only 6 have both \lya\ and H$\alpha$ detection (Hayes et al. 2010). So the escape fraction at z$\sim$ 2.25 estimated by Hayes et al. (2010) is a global \lya\ escape fraction for all galaxies at that redshift. 
The narrowband surveys for high-z LAEs can be designed to use the same criteria, e.g., the threshold equivalent width of 
EW$_{rest}(Ly\alpha)$ $\geq$ 20 \AA\ and the \lya\ line flux limit estimated from narrowband and broadband images, 
however, we should note that the \lya\ escape fraction is dependent on the choice of the threshold equivalent width, and the narrowband exposure depth. 
Shallower narrowband surveys tend to select bright LAEs, which may have smaller \lya\ EWs (e.g., Ando et al. 2006), and a decrease in the value of
threshold equivalent width will contain more objects with little-to-no \lya\ fluxes as well as interlopers. 


X-ray photons, especially hard X-ray photons, can give an extinction-free measurement of 
star formation in galaxies.  First, 
the X-ray emission of high-redshift 
star-forming galaxies are mainly from supernovae (SNe), hot interstellar gas (i.e., $T >
10^{6-7}$ K), and high-mass X-ray binaries (HMXBs), all of which trace
star formation ongoing in the last few million years.  
Nandra et al. 2002, Grimm et al. 2003, Ranalli et al. 2003, and Gilfanov, Grimm \& Sunyaev 
(2004) have shown a linear correlation between L$_X$ and SFR.   Calibration with the 
total SFR from far-infrared (FIR) and ultraviolet (UV) bands for local and high-z (z $\sim$ 1 and z $\sim$ 3) 
star-forming galaxies (in the Chandra Deep Field North and South, CDF-N and CDF-S, where the 
deepest X-ray observations exist) show relations consistent with Ranalli et al. (2003)\footnote{Ranalli et al. (2003) had calibrated the relations between X-ray luminosities and radio/FIR luminosities, and converted to global SFR (refereed to stars with $M$ $>$ 5 M$_\sun$ and Salpeter's Initial Mass Function (IMF)) following Condon (1992, radio luminosities) and Kennicutt (1998, FIR luminosities). }:
 \begin{eqnarray}
SFR_X &  = & 2.2 \times 10^{-40} (L_{0.5-2} / \hbox{erg}\, \hbox{s}^{-1}) \mbox{       M}_\sun \mbox{ yr}^{-1},  \nonumber\\
SFR_X &  = & 2.0 \times 10^{-40} (L_{2-10}  / \hbox{erg}\, \hbox{s}^{-1})\mbox{       M}_\sun \mbox{ yr}^{-1}.
\end{eqnarray}

Second, X-rays penetrate a typically dusty interstellar medium much more
easily than UV light.  For example, a column density of $> 10^{22} $cm$^{-2}$ is
required to attenuate 2 keV X-rays by a factor of 2.   The corresponding reddening would be
$E(B-V)=1.7$ mag, corresponding to $A_V \approx 5$ mag and $A_{UV} \ga 10$ mag, or a
factor of $10^4$, based on ratios of dust to gas from either our Galaxy ($E(B-V) = 1.7 \times 
10^{-22}$ mag cm$^2$ atoms$^{-1}$) or from a set of three gravitationally lensed galaxies at high
redshift ($E(B-V)/N_H$ = (1.4 $\pm$ 0.5) $\times$ 10$^{-22}$ 
mag cm$^2$ atoms$^{-1}$; Dai et al 2006).

In this paper, we present an independent analysis of the \lya\ escape fraction at $z$ $\sim$ 0.3, $z$ $\sim$ 1, and 2 $<$ $z$ $<$ 6.5, using X-ray emission 
as a tracer of the intrinsic SFR. We use the new 4 Msec {\it Chandra} X-ray image in the CDF-S, which include
LAEs at z $\sim$ 0.3 and 1.0 selected by GALEX, and ground-based narrowband selected LAEs at z = 2.1, 3.1, 3.15, 4.5, 5.7, and 6.5. By comparing 
the ratio of derived SFRs from observed \lya\ line flux and X-ray, we will perform an independent measurement
of the \lya\ escape fraction for LAEs. The optical samples and X-ray data are 
presented in \S 2, and the X-ray detection and stacking results on LAEs are presented in \S 3. The results 
and discussion on X-ray constrained \lya\ escape fraction are presented in \S 4 and \S 5, respectively. 


\section{OPTICAL AND X-RAY DATA}
In the Chandra Deep Field South region, samples of \lya\ emitters have been observed at various redshifts,
including \lya\ emitters at z = 0.195-0.44 and z = 0.65-1.25 through GALEX grism spectroscopy (Deharveng et al. 2008, Cowie et al. 2010, Cowie et al. 2011),
and ground-based narrowband imaging selected \lya\ emitters at z = 2.1 (Guaita et al. 2010),  z = 3.1 (Gronwall et al. 2007, Ciardullo et al. 2011 in prep.), z = 3.15 
(Nilsson et al. 2007), z = 4.5 (Finkelstein et al. 2009), z = 5.7 (Wang et al 2005) and z = 6.5 
(Rhoads et al in prep.). The GALEX LAEs were confirmed by ground based optical spectroscopy (Cowie et al. 2010, 2011). 
At higher redshift (z $>$ 2), all the narrowband selected samples typically have a spectroscopic
 confirmation fraction greater than 70\% (Dawson et al. 2007, Gawiser et al. 2007, Wang et al. 2009). 
In this paper we focus on LAEs covered in the CDF-S proper region. 
The number of LAEs at each redshift and their stacked X-ray properties are presented in Tables 1 and 2.

The 4 Ms  \textit{Chandra} Advanced CCD Imaging Spectrometer (ACIS) exposure of the 
CDF-S is composed of 52 individual ACIS-I observations, 9 of which were obtained in 2000,
 12 from September to November 2007, and 31 from March to July 2010. The event-2 file and 
 exposure-map files are available at the {\it Chandra} website (http://cxc.harvard.edu/cdo/cdfs.html). 
 To search for potential X-ray counterparts for our LAEs, we use the 4-Ms CDF-S catalog published
 by Xue et al. (2011) for the following source cross-match and source-masking processes. 
 

\section{X-RAY INDIVIDUAL DETECTION AND STACKING PROCEDURE}

\subsection{Individual detections in Chandra images}

We search for X-ray counterparts for individual LAEs within a 2$\arcsec$ radius.  Among LAEs selected
in the 4 Msec CDF-S region using the GALEX FUV channel (which covers \lya\ at $0.195< z < 0.44$), 
3 of 6 (found at $0.2<z<0.37$) 
are detected in X-rays.   These three have relatively modest luminosities 
L$_{2-10 keV}$ $<$ 10$^{42}$ ergs s$^{-1}$, and have been classified as galaxies from their optical spectra (Cowie et al. 2010, 2011). Their UV and X-ray properties are presented in Table \ref{z03lae}.  Note that all three detected galaxies show a soft-band X-ray detection, while their hard-band luminosities
are upper limits, but remain consistent with their soft band luminosities under the $\Gamma$ = 2 assumption.   For \lya\ galaxies selected in the GALEX NUV channel (which covers \lya\ at
$0.65<z<1.2$), the corresponding number is 4 X-ray detections out of 5 objects in the 4 Msec 
Chandra coverage.  These  four objects have 
X-ray luminosities L$_{2-10 keV}$ $>$ 6$\times$10$^{43}$ ergs s$^{-1}$, and 
can be classified as AGNs.

X-ray counterparts have previously been detected for a small fraction ($<$ 5\%)  of the LAEs at higher redshifts. 
Guaita et al. (2010) detected X-rays from 10 of their 216 $z$ = 2.1 LAEs in the ECDF-S region, 
using the 2Msec {\it Chandra} image for those LAEs in the central CDF-S (Luo et al. 2008), and the 250 ksec {\it Chandra} image
(Lehmer et al. 2005) for those in the wider ECDFS. Gronwall et al. (2007) and Ciardullo et al (in prep.)
selected 278 LAE candidates at z =  3.1, and only 5 were X-ray detected. At z = 4.5, only 1 of 113 LAE candidates show X-ray detection 
in ECDF-S field (Zheng et al. 2010) and is classified as an AGN at the same redshift, and no X-ray detections were 
found for 24 LAE candidates at z = 3.15 (Nilsson et al. 2007),  25 LAE candidates at z = 5.7 (Wang et al. 2005), or 11 LAE candidates at z = 6.5 (Rhoads et al. in prep.). 
In this paper, these previous X-ray detections are excluded, and we find no new X-ray detection with the 4 Ms Chandra exposure. 

We calculate the X-ray signal-to-noise ratio for the X-ray non-detected LAEs, all of which show S/N $<$ 3.
The S/N ratios were calculated as $S/N = S/(\sqrt{T+0.75}+1)$  (Gehrels 1986), 
where $S$ and $T$ are the net counts and total counts extracted from their 50\%  Chandra ACIS PSF
circles\footnote{ The background counts (${\rm B = T - S}$) were estimated from an annulus with 1.2R$_{95\%PSF}$ $<$ R $<$ 2.4R$_{95\%PSF}$ (after masking out nearby
X-ray sources). } in 0.5-2, 2-7 and 0.5-7 keV band respectively. 
To convert from X-ray counts to fluxes, we have assumed a powerlaw spectrum with photon index of  $\Gamma$ = 2 (except where explicitly stated otherwise), which generally represents the X-ray spectra of both starburst galaxies (e.g., Lehmer et al. 2008) and type 1 AGNs.
When converted from PSF-corrected count-rate  to flux, the full and hard bands 
were extrapolated to the standard upper limit of 10 keV.  All X-ray fluxes and luminosities presented in this paper have been corrected for Galactic absorption (Dickey \& Lockman 1990).  The 4 Ms CDF-S data reaches on-axis sensitivity limits of X-ray luminosity Lg$_{10}(L_{2-10})$ $\approx$ [39.6, 40.9, 41.5, 41.9, 41.9, 42.3, 42.5, 42.6] at z = [0.3, 1.0, 2.1, 3.1, 3.2, 4.5, 5.7, 6.5] (with the on-axis soft band flux limit of $f_{0.5-2 keV}$ $\approx$ 9.1 $\times$ 10$^{-18}$ erg cm$^{-2}$ s$^{-1}$, Xue et al. 2011), which means that we may find all the AGNs (including type 2 AGNs) hidden in the z$\sim$0.3 and z$\sim$1 LAEs located in the CDF-S region.

\subsection{Stacking analysis} \label{sec:stacka}

We stacked the X-ray data at the positions of the non-detections at each redshift, following Zheng et al. (2010).
After masking out the X-ray detected sources with 95\% PSF circles, we
co-added the net and background counts of the X-ray non-detected LAEs in each sample.
The stacked net and background counts in the soft, hard and 0.5-7 keV band, 
as well as the summed effective Chandra exposure time are presented in Table \ref{stkdata}. 
We used the 95\% PSF region for bright source extraction, and 50\% PSF region for upper limit 
estimate\footnote{PSF radii at various flux fractions were based on a polynomial approximation 
of the off-axis ACIS-I PSF, as discussed in http://cxc.harvard.edu/chandra-users/0192.html.}.
The results are about 0.2 dex lower than upper limit from 95\% PSF extraction.

\begin{figure}
\plotone{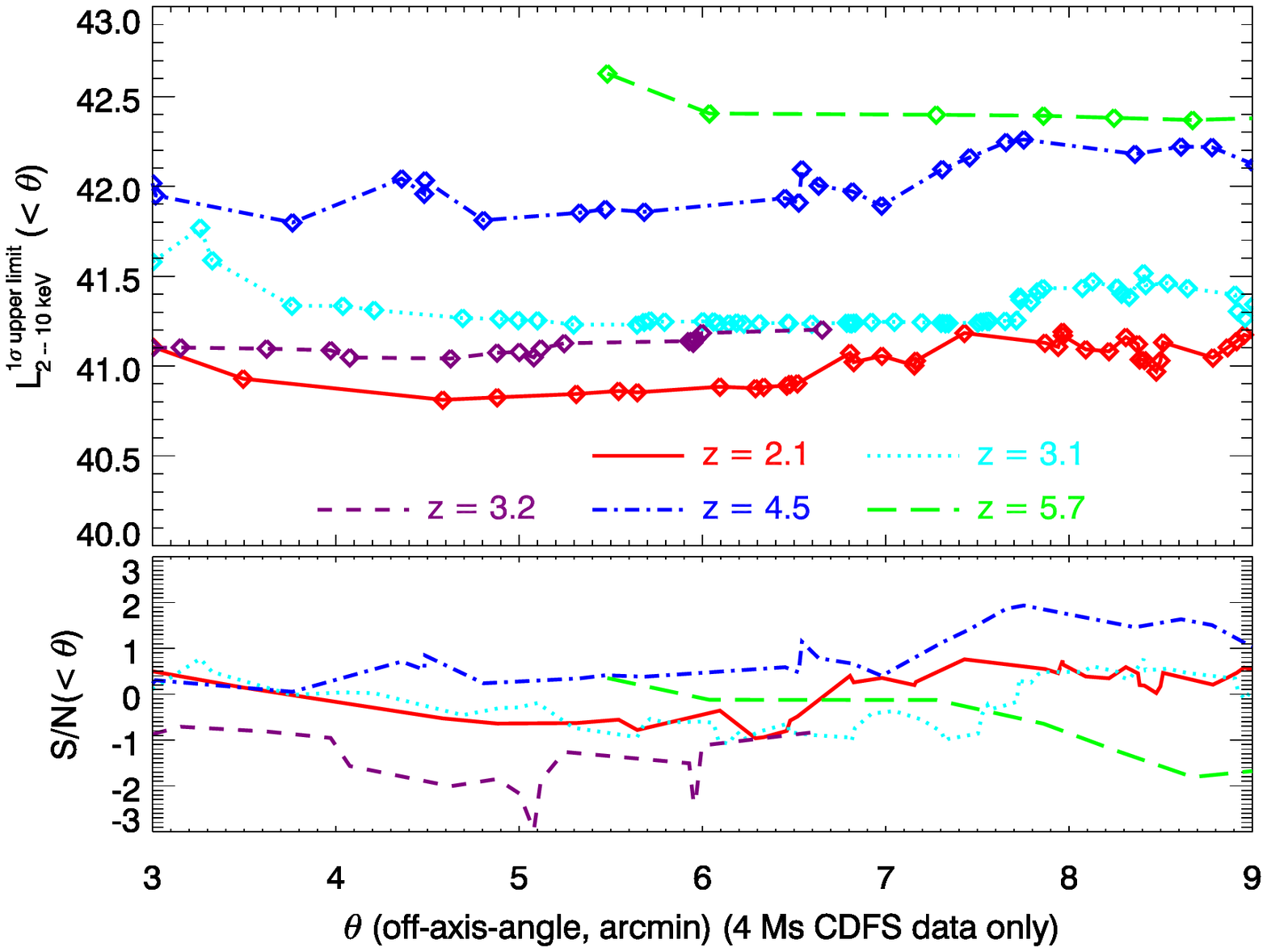} 
\caption{ The X-ray stacking results for LAEs located within off-axis-angle $< \theta$ in the 4 Ms CDF-S data. The lower plot shows the
signal-to-noise ratio of the stacked signal, and the upper plot shows the 1-$\sigma$ upper limit on L$_{2-10 keV}$ from the stacked signal.
It can be seen that the upper limits on L$_{2-10 keV}$ tend to increase when stacking LAEs located with $\theta$ $\gtrsim$ 6.5 arcmin. 
In this work the upper limit on f$_X$ and L$_X$ are extracted from stacking LAEs located within an off-axis angle $\theta$ $<$ 6 arcmin.
} 
\label{sntheta}
\end{figure}

In Figure \ref{sntheta} we plot the 1 $\sigma$ upper limits  (upper plot)  on the mean X-ray luminosity  
for each studied redshift in the selection region (parameterized by the maximum off-axis angle $\theta$ for inclusion in the sample),
and the S/N ratios of the stacked signals are also plotted in the lower part. 
From Figure \ref{sntheta} we can see than the mean luminosity can be better constrained by excluding LAEs with larger off-axis angles. 
This is mainly because the Chandra ACIS has much larger PSF and much lower collection area at larger off-axis angles, 
so that including those LAEs with large off-axis angles would bring strong fluctuations to the signal without necessarily increasing the S/N. 
In the following study, we exclude those LAEs with ACIS off-axis angle above  6$\arcmin$, and the number of LAEs used for stacking and their stacked results at 
relative redshifts are presented in Table \ref{stkdata}.


\section {RESULTS AND DISCUSSION}

\subsection{SFR-Luminosity Correlation}

We adopted the relations established at z = 0 to convert luminosities to SFRs for our LAE samples: 
Kennicutt (1998) for conversion of the \lya\ luminosity (assuming Case-B conditions and Salpeter's IMF with mass limits 0.1 and 100 $M_\sun$), 
and Ranalli et al. (2003) for the X-ray luminosities:
\begin{eqnarray}
SFR_{Ly\alpha} &  = & 9.1 \times 10^{-43} L_{Ly\alpha} \mbox{       M}_\sun \mbox{ yr}^{-1},  \nonumber\\
SFR_X &  = & 2.0 \times 10^{-40} L_{2-10} \mbox{       M}_\sun \mbox{ yr}^{-1}. 
\end{eqnarray}
Note that these relations might give uncertainties as well as the factor of $\sim$2 dispersion (e.g., dust extinction,
burst age and IMF) when applied to individual sources. However, previous X-ray stacking works (e.g., Nandra et al. 2002, 
Seibert et al. 2002, Lehmer et al. 2005, and Reddy \& Steidel 2004) have proved reasonable L$_X$-SFR$_X$ connection on local 
starburst galaxies and 1 $\leq$ z $\leq$ 3 LBGs. 

Table \ref{z03lae} shows the SFR estimates based on the \lya\ (``SFR$_{Ly\alpha}$'') luminosity and 2-10 keV (``SFR$_X$'') 
luminosity for three z$\sim$0.3 GALEX selected LAEs.  Table \ref{stkdata} shows the SFR$_X$ estimates from the stacking 
signal of the X-ray undetected LAEs at z $\sim$ 0.3, 1.0, 2.1, 3.1, 3.2, 4.5, 5.7 and 6.5. It can be seen that at z $\geq$ 4.5, the 
SFR$_X$ limits are weaker due to a larger luminosity distance, and comparatively smaller \lya\ sample size.

Note that low-mass X-ray binaries (LMXBs) 
can also contribute to X-ray emission from galaxies, but they have longer 
evolutionary time scales (on the order of the Hubble time), and therefore track the integrated 
star-formation history of galaxies (i.e., the total stellar mass). 
Where LMXBs are important, e.g. for 
nearby normal galaxies (Colbert et al. 2004) and luminous infrared galaxies 
(Lehmer et al. 2010), the X-ray luminosity can be fitted as a function of both galaxy 
galaxy stellar mass and SFR:
\begin{equation}
L_{2-10 keV} = \alpha\times M_* + \beta \times SFR_X ~~.
\end{equation}
Here Lx , M$_*$, and SFR$_X$ have units of ergs s$^{-1}$ , M$_{\odot}$ , and M$_{\odot}$ yr$^{-1}$, 
respectively, and constants $\alpha/(10^{29}$ ergs s$^{-1}$ M$_{\odot}^{-1})$ = 1.51 and 0.91, and 
$\beta/(10^{39}$ ergs s$^{-1}$ (M$_{\odot}$ yr$^{-1}$)$^{-1})$ = 0.81 and 1.62 for Colbert et al. (2004) and Lehmer et al. (2010), respectively. 
The differences in the parameters might be introduced by the high obscuration in luminous infrared galaxies. 
We can see from these relations, the contribution of X-ray emission from LMXBs can be ignored in 
actively star-forming galaxies, which often 
have low dust extinction (0$\lesssim$ E(B-V) $\lesssim$ 3), lower stellar mass (M$_*$ $\lesssim$ 10$^{9}$ M$_{\odot}$), and 
more active star formation (SFR$_X$ $\gtrsim$ 1 M$_{\odot}$ yr$^{-1}$) than local galaxies (e.g., Guaita et al. 2011).  In terms of the specific star formation rate (SSFR), which is defined
as the ratio of star formation rate to stellar mass,  we can neglect the LMXB term whenever 
$\beta \times SFR_X \gg  \alpha\times M_*$, so that $SSFR \gg \alpha/\beta \sim 10^{-9} \hbox{yr}^{-1}$,
i.e., whenever the typical stellar ages are much below 1 Gyr.


The SFR$_X$ of high redshift star-forming galaxies was measured by Nandra et al. (2002), who detected a 
stacked X-ray signal of $z \approx$ 3 Lyman Break Galaxies (LBGs) in the Hubble Deep Field North. 
The average X-ray luminosity of $z \approx$ 3 LBGs is L$_{2-10}$ = 3.4 $\times$ 10$^{41}$ ergs s$^{-1}$ 
(6-$\sigma$ significance), implying a SFR$_X$ = 64$\pm$13 M$_\sun$ $yr^{-1}$, in excellent agreement with the extinction-corrected UV estimates.
 In the same field, Laird et al. (2006) found the average SFR$_X$ of 42.4$\pm$7.8 M$_{\odot}$ yr$^{-1}$ for z $\sim$ 3 LBGs.  
 Additionally, Lehmer et al. (2005) reported the average SFR$_X$ of $\sim$30 M$_{\odot}$ yr$^{-1}$ for z $\sim$ 3 LBGs in CDF-S (old 1 Ms {\it Chandra} exposure). 
 Lehmer et al. also stacked LBGs in CDF-S at z$\sim$4, 5, and 6, and did not obtain significant detections ($<$3 $\sigma$), 
 deriving rest-frame 2.0-8.0 keV luminosity upper limits (3 $\sigma$) of 0.9, 2.8, and 7.1 $\times$ 10$^{41}$ ergs s$^{-1}$, 
 corresponding to SFR$_X$ upper limits of 18, 56 and 142 M$_{\odot}$ yr$^{-1}$, respectively. Note also that a $\sim$3 $\sigma$ 
 stacking signal of the optically bright subset (brightest 25\%) of LBGs at z$\sim$4 was detected, 
 corresponding to an average SFR$_X$ of $\sim$28 M$_{\odot}$ yr$^{-1}$. Reddy \& Steidel (2004) examined the stacked radio and X-ray emission 
 from UV-selected spectroscopically confirmed galaxies in the redshift range 1.5 $\lesssim z \lesssim$ 3.0. 
 Their sample showed SFR$_X$ $\sim$ 50 M$_{\odot}$ yr$^{-1}$, and  found a consistent SFR$_X$/SFR$_{UV, uncorr}$ $\sim$ 4.5-5.0 
 for galaxies over the redshift range 1.5 $\lesssim z \lesssim$ 3.0.  These studies demonstrate 
 the value of stacking the deepest X-ray observations to measure star formation activity, with little sensitivity to dust.

The LAEs' X-ray emission should be directly connected to the intrinsic SFR, since LAEs are thought to be less massive and much younger than LBGs at high-redshift 
(e.g., Venemans et al 2005; Pirzkal et al 2007; Finkelstein et al 2008, 2009c), and the AGN fraction in LAEs is also very small ($<$ 5\%, e.g., Zheng et al. 2010).  An X-ray
detection could give us a more accurate unbiased SFR estimate, or more properly an
upper limit, since faint AGN may contribute to the X-ray flux. The first X-ray observations of high--redshift LAEs were 
presented in Malhotra et al. (2003) and Wang et al. (2004) at z$\sim$4.5 with two 170 ks {\it Chandra} exposures.
No individual LAEs were detected, and a 3-$\sigma$ upper limit on the X--ray
 luminosity (L$_{2-8 keV}$ $<$ 2.8 $\times$ 10$^{42}$ ergs s$^{-1}$) was derived by an X-ray stacking method (Wang et al. 2004). 
From a stacking analysis of the non-detected LAEs in the much deeper 2 Ms CDF-S field and a larger 250 ks ECDFS field, 
 Guaita et al. (2010), Gronwall et al. (2007) and Zheng et al. (2010) found a smaller 
3-$\sigma$ upper limit on the luminosity of $\sim$1.9 $\times$ 10$^{41}$ ergs s$^{-1}$, 3.1 $\times$ 10$^{41}$ 
ergs s$^{-1}$ ,2.4 $\times$ 10$^{42}$ ergs s$^{-1}$ at z = 2.1, z = 3.1 and z = 4.5, respectively.  These imply 
upper limits of unobscured average SFR$_X$ $<$  43, 70 and 290 M$_{\odot}$ yr$^{-1}$, respectively 
. The above results are not surprising, since
we would expect that LAEs have lower SFR rates than LBGs.

\subsection{f$_{esc}^{Lya}$ from X-rays}

The average \lyaesc\ calculated using SFR$_{Ly\alpha}$ and SFR$_X$ for LAEs at different redshifts are 
presented in Table \ref{xsfr}. The sample selection limits are also presented in Table \ref{xsfr}. We should
note that the LAEs at z$\sim$ 0.3 and 1 are selected among the GALEX FUV and NUV band drop-out galaxies with $EW(Ly\alpha)$ $>$ 20\AA\
from their follow-up GALEX spectra.  So the LAE samples at z $\lesssim$ 1 are quite different compared to the LAE samples at z$>$ 2, which 
are selected from narrowband excess over broadband with nearly same criteria on the equivalent width $EW(Ly\alpha)$ (estimated from the broad band to narrowband ratio). 
In the following discussion, we treat them separately. 
We only take the soft band upper limits for z$>$ 2 LAEs,
 because soft band flux are more sensitive than the total band and hard band, and at z $>$ 2, 
 the observed soft band X-ray photons are closer to rest-frame hard photons, 
 and therefore more robust to a change in photon index $\Gamma $ (see Figure 2 of Wang et al. 2007) assumed when converting X-ray count-rate to flux. 

At $z \sim$ 0.3, we only have L$_{0.5-2 keV}$ for the three \lya\ selected galaxies. Although they are not detected in the hard X-ray band, their 
3 $\sigma$ upper limit on L$_{2-10 keV}$ show good consistency with their L$_{0.5-2 keV}$ (See table 1). 
Here we use the SFR$_X$-L$_{0.5-2 keV}$ relation from Ranalli et al. (2003), since 
our detections are in the soft band.  We find SFR$_X$ $\sim$ 1, 11, and 22 M$_\sun$ yr$^{-1}$, implying 
\lya\ escape fractions of 22, 3, and 9\%, respectively. 
Finkelstein et al. (2009b) also derived the 
 dust extinction of A$_{1200}$ = 0.9$\sim$2 from the SED fitting, corresponding to $f^{Esc}_{Ly\alpha/UV-corr}$ 
 = SFR$_{Ly\alpha}$/(SFR$_{UV}/10^{-A_{1200}/2.5}$) = 2-5\% (see table 3), which is systematically lower than the \lyaesc\ from X-ray.  X-ray stacking of the undetected z $\sim$ 0.3  LAEs shows a 3 $\sigma$ upper limit of SFR$_X(3\sigma)$ $\sim$ 
3.2 M$_\sun$ yr$^{-1}$, implying an average \lya\ escape fraction of $>$28\% at 3 $\sigma$ significance level. 
These X-ray constraints may tell us that at z$\sim$ 0.3, the \lya\ escape fraction is substantially
larger than we thought from the UV or H$\alpha$ view. 

There are only 2 star-formation dominated LAE galaxies at z$\sim$ 1 in the ECDFS
region (excluding the four AGN with individual X-ray detections), 
so at z$\sim$ 1 the escape fraction is poorly constrained (with \lyaesc($>1 \sigma$) $>$ 7\%).
 
\begin{figure}
\plotone{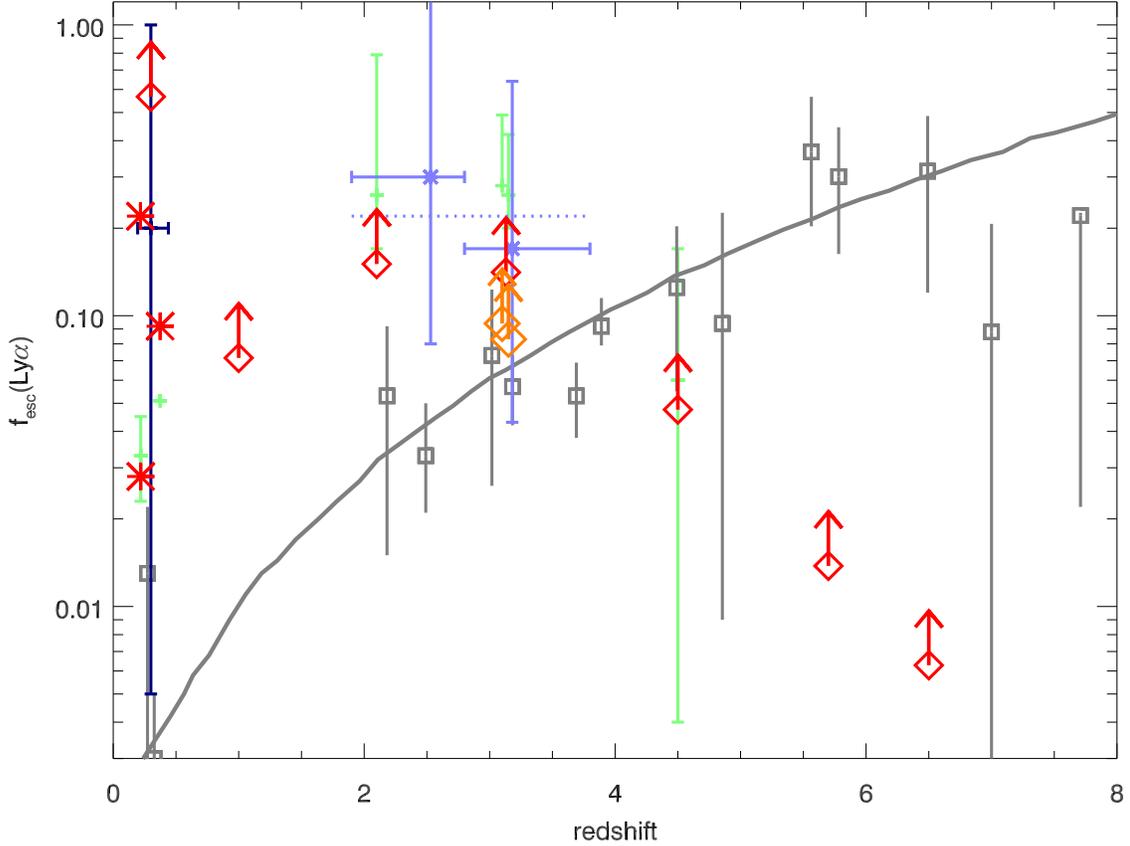} 
\caption{ Lya escape fraction estimated from Optical (navy cross: \lyaesc\ of z $\sim$ 0.3 LAEs from Atek et al. 2009; 
light blue crosses and dotted lines: \lyaesc\ from blank field spectroscopic survey of 98 LAEs at 1.9 $<$ z $<$ 3.8 from 
Blanc et al. 2011; green crosses: \lyaesc\ from SED fitting results of our sample, grey squares: global \lyaesc\ extracted from figure 1 of Hayes et al. 2011) 
compared with estimated from X-rays 
(red $*$: three LAEs with X-ray detections at z $\sim$ 0.3; red diamond limits: \lyaesc\ 1-$\sigma$ lower limits for LAEs at z = [0.3, 1.0, 2.1, 3.1, 4.5, 5.7, 6.5]). Note that at z$>$4 and z $\sim$ 1, there are not enough data to constrain on X-ray derived \lyaesc.  The orange diamonds are the separate 1-$\sigma$ upper limit at z $\sim$ 3.1 and 3.2. The grey line  is the best-fit constraint on the redshift evolution of global $f_{esc}^{Ly\alpha}$ derived by Hayes et al. 2011. 
} 
\label{lyafrac}
\end{figure}

At $z$ $>$ 2, we do not obtain an X-ray detection after stacking the 69 star-formation dominated LAEs in 
the central CDF-S field, or after stacking the 351 LAEs in the ECDFS field (see Table \ref{stkdata}).   
We get the 1-$\sigma$ upper limits (same as below) on the intrinsic (dust-free) SFRs as SFR$_X$ $<$ [14, 28, 28, 139, 440, 876] $
M_\sun$ yr$^{-1}$ for LAEs at z = [2.1, 3.1, 3.2, 4.5, 5.7, 6.5]. The observed SFRs from the \lya\ line are about 1.9, 2.6 and 2.3 
M$_\sun$ yr$^{-1}$ for z = 2.1, 3.1 and 3.2 LAEs on average, and $\sim$ 5 M$_\sun$ yr$^{-1}$ for z $>$ 4 LAEs. The 
ratio SFR$_{Ly\alpha}$ / SFR$_X$ measures the \lya\ escape fraction: 
\lyaesc$_{/X}$= SFR$_{Ly\alpha}$ / SFR$_X$ (see Table 3). This is plotted in figure \ref{lyafrac}, showing the constraints on \lyaesc\ as 
a function of redshift. At z = 2.1, the \lya\ escape fraction of LAEs is above 14\%. At z = 3.1 and 3.2, the f$_{esc}^{Lya}$ 
is greater than 8\% and 9\%, separately. Combining the two samples at z$\sim$ 3, we increase the upper limit to f$_{esc}^{Lya}$ $>$ 14\%, 
the same as the upper limit at z = 2.1. At higher redshift, the limits are weaker due to a larger luminosity distance, 
and comparatively smaller \lya\ sample sizes.

If the \lya\ emitters don't evolve from z = 3.2 to z = 2.1, then we can combine the samples in this redshift range to obtain a 
more robust limit on the escape fraction. A 1-$\sigma$ upper limit
 on soft band flux is derived as f$_{0.5-2 keV}$ $<$ 8.5$\times$ 10$^{-19}$ \fluxunit. This 
 implies an SFR$_{Ly\alpha}/SFR_X >$ 14\%$\sim$28\% (due to the different SFR$_{Ly\alpha}$ average value)  
 during the redshift range 2.1 $\le$ $z$ $\le$ 3.2. This value is consistent with the median value of 
$f^{Esc}_{Ly\alpha/UV\_corr}$ $\sim$ 29\% from blank fields IFU spectroscopically-selected LAEs at 1.9 $< z <$ 3.8 
 (Blanc et al. 2011), but larger than the value of $f^{Esc}_{Ly\alpha}\sim$ 2, 5-10\% in some theoretical models of LAEs 
 (e.g., Le Delliou et al. 2005, Shimizu et al. 2011).


\subsection{$f^{Esc}_{Ly\alpha}$ with 4 Msec CDF-S data}

There is no X-ray detection for \lya\ emitters coadded at any individual redshift bin (z $>$ 2). 
By coadding the 53 \lya\ emitters at 2.1 $\leq z \leq$ 3.2, we reach an X-ray flux limit of 
8.5$\times$10$^{-19}$ ergs cm$^{-2}$ s$^{-1}$ (1 $\sigma$), but still do not detect X-ray photons from these galaxies. 
This means that the star-formation rate is truly low, as indicated by the \lya\ line strength. To increase our 
sensitivity to detect the star formation rate of LAEs, which is relatively low when compared to continuum-selected 
star-forming galaxies at comparable redshifts, we coadd all the undetected \lya\ emitters between redshifts 
2 and 6.5. The average x-ray emission is still undetected at a 1-sigma flux level of 7.6$\times$10$^{-19}$ 
ergs cm$^{-2}$ s$^{-1}$. To convert the flux into luminosity and hence SFR, we need to model the redshift distribution of the sources.
 
\begin{figure}
\plotone{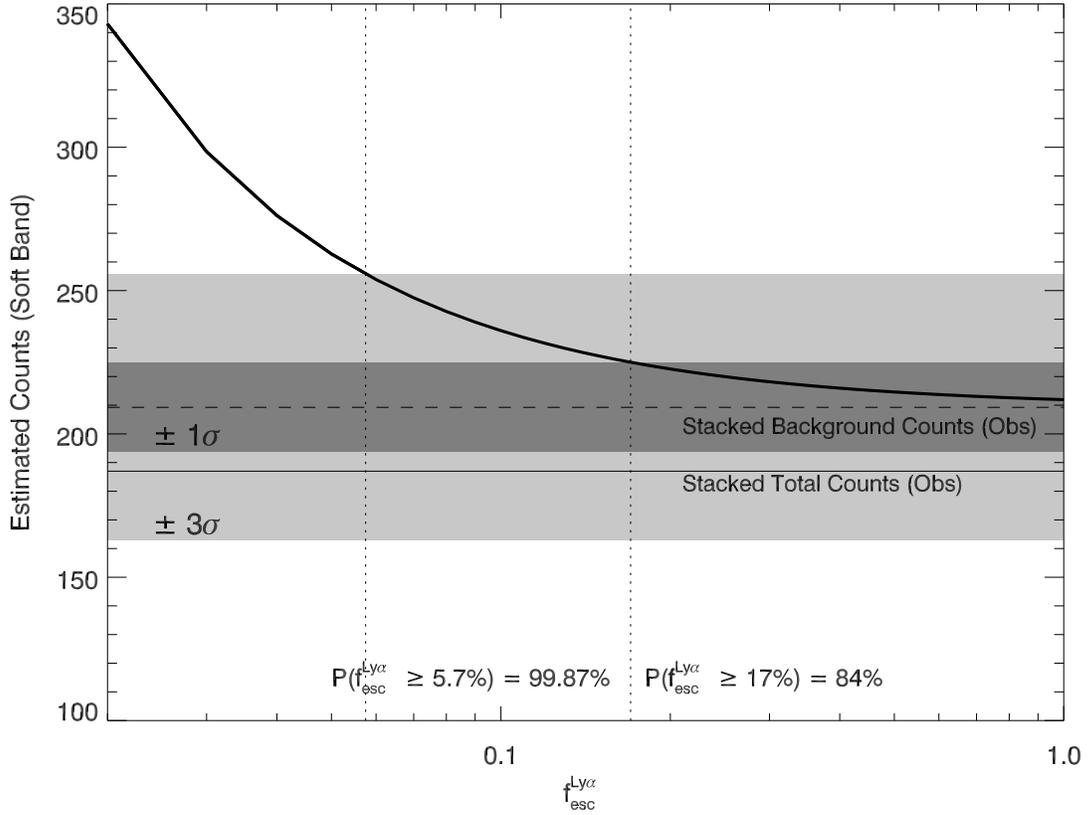}
\caption{ The estimated X-ray total counts as a function of  $f_{esc}^{Ly\alpha}$ for all LAE candidates in CDF-S at 2 $<  z < $7. Note that the observed total count (the solid horizontal line) in soft X-ray band is lower than the background count (the dashed horizontal line). The two dotted vertical lines imply where the estimated signal reach 1-$\sigma$ and 3-$\sigma$ level. 
} 
\label{estcts}
\end{figure}

Let us, instead, predict the expected X-ray counts on this average image of  \lya\ galaxies. Assuming the  
\lya\ escape fraction for LAEs is $f^{Esc}_{Ly\alpha}$, then SFR$_X$ = SFR$_{Ly\alpha}$/($f^{Esc}_{Ly\alpha}$)=  (L$_X / 5 \times 10^{39} \hbox{erg} \, \hbox{s}^{-1}) M_\odot \hbox{yr}^{-1}$, 
we get 
\begin{equation}
L_{2-10 keV,\mbox{ rest}} =  5\times 10^{39} \hbox{erg}\, \hbox{s}^{-1} \times \frac{SFR_{Ly\alpha} / M_\odot \, \hbox{yr}^{-1}}{2.0\times f^{Esc}_{Ly\alpha}} \times 10^{40}\mbox{ ergs s}^{-1}.
\end{equation}
If we assume that all LAEs have an effective X-ray photon index of $\Gamma$ = 2, then with the
X-ray count-rate to flux conversion of 6.64 $\times$ 10$^{-12}$ erg/cm$^2$ at soft band and SFR$_{Ly\alpha}$ from the table \ref{xsfr}, we estimate that the 
expected number of  X-ray photons at $\sim$4 Ms CDF-S for LAEs should be $ (\frac{SFR_{Ly\alpha}}{ f^{Esc}_{Ly\alpha} \times 100}\times[8.0, 3.0, 3.0, 1.3, 0.7, 0.5])$ 
for LAEs at $z = [2.1, 3.1, 3.2, 4.5, 5.7, 6.5]$. This means that we should observe $\sim$ 3 soft X-ray 
photons per z = 2.1 LAE galaxy in the 4 Ms CDF-S field when $f^{Esc}_{Ly\alpha}$ $\sim$ 5\%,  and the 
observed expected soft X-ray photons decrease to $\sim$ 1 when $f^{Ly\alpha}_{esc}$ increases to 15\%. 
Since the background X-ray photons per point source on the ACIS-I CCDs varies from position to position, 
we take the background value extracted from each LAE candidate, and add the estimated X-ray photons 
from their corresponding SFR$_{Ly\alpha}$ to analyse the probability on $f^{Esc}_{Ly\alpha}$. By coadding 
all LAEs between redshift z = 2 and 6.5, the estimated counts are plotted in Figure \ref{estcts} as a function 
of  $f^{Esc}_{Ly\alpha}$. The signal should be S/N $ > $ 3 when $f^{Esc}_{Ly\alpha}$ $<$ 5.7\%, and S/N $ > 1$ 
when $f^{Esc}_{Ly\alpha}$ $<$ 17\%. So we can reject the value of 5\% at 99.87\% confidence 
level, and report that the real value of  $f^{Esc}_{Ly\alpha}$ $>$ 17\% in 84\% confidence level. 

\subsection{$f^{Esc}_{Ly\alpha}$(X-ray) vs. $f^{Esc}_{Ly\alpha}$(UV/Opt.) }

The escape fraction of LAEs has also been discussed using other tracers of the
total SFR.   Atek et al. (2009) studied z = 0.3 LAEs by using the 
extinction-corrected H$\alpha$ to \lya\ ratio, which has a  ranges of $f^{Esc}_{Ly\alpha/H\alpha}$$\sim$ 0.5\% to  100\% and a median value of $\sim$ 20\%.  
Blanc et al. (2011) studied 1.9$ < z <$3.8 LAEs through extinction-corrected UV to \lya\ ratio 
and found an  average $f^{Esc}_{Ly\alpha/UV\_corr}$ $\sim$ 29\%. 
The f$_{esc}^{Lya}$ for LAEs estimated from the ratio of SFR$_{Ly\alpha}$ and dust-corrected 
SFR$_{UV}$, $f^{Esc}_{Ly\alpha/UV\_corr}$ = SFR$_{Ly\alpha}$/(SFR$_{UV}/10^{-A_{1200}/2.5}$),  
also exist for our sample. Guaita et al. (2011), Gawiser et al. (2007) and Nilsson et al. (2007) did the stacked SED fitting on the samples at z = 2.1 $\sim$ 3.2. At z$\sim$ 0.3 and 4.5, 
Finkelstein et al. (2008, 2010) did  individual SED fitting for LAEs with existing {\it Hubble} and {\it Spitzer} observations.  
The dust properties of the SED fitting results at different redshifts are converted to A$_{1200}$ (see in table 3) 
using Calzetti et al. (2000). The $f^{Esc}_{Ly\alpha/UV\_corr}$ from dust-corrected UV to \lya\ ratio for our sample show no evolution from z$\sim$2 to z$\sim$3.2, 
as $f^{Esc}_{Ly\alpha/UV\_corr}$ $\sim$ 26\%, and is consistent with  $\sim$ 29\% of Blanc et al. 2011. At z $\sim$ 4.5, the SED fitting results
might be affected by the poor spatial resolution of {\it Spitzer}.  As an 
independent estimate on \lyaesc\ from X-ray, our  \lyaesc\ at z = 0.3 are located at the low end of 
Atek et al. (2009), while consistent at 1.9 $< z <$ 3.8 with Blanc et al. 2011 (see figure \ref{lyafrac}), and the
$f^{Esc}_{Ly\alpha/UV\_corr}$ from the SED fitting results at  at z = 2.1, 3.1,  3.2 and 4.5.
The LAEs at $z \sim$ 0.3 seems different compared to high-redshift LAEs, as they are more 
AGN contaminated (AGN fraction of $\sim$15\%--40\%,  e.g., Scarlata et al. 2009; Cowie et al. 2010; 
Finkelstein et al. 2009a) and more massive (Finkelstein et al. 2009b). At high-redshift, the AGN fraction in LAEs is 
very low (AGN fraction $\lesssim$ 5\%, Zheng et al. 2010 and references there in), and AGNs are relatively easy to detect in 
this deepest {\it Chandra} field, so the lower limit of f$_{esc}^{Lya}$ should be very robust. Our X-ray constraints on the $f^{Esc}_{Ly\alpha}$, 
as well as the $f^{Esc}_{Ly\alpha/UV\_corr}$ estimated from SED fitting, show that the \lya\ escape fraction need not evolve during the redshift 2.1 $\leq$ $z$ $\leq$ 3.2.

Hayes et al. (2011) and Blanc et al. (2011) also reported the global evolution of f$_{esc}^{Lya}$, which is defined 
as the ratio of integrated \lya\ luminosity functions from LAEs and the global extinction-corrected SFR density at different redshift. 
The global extinction-corrected SFR densities were integrated from H$\alpha$ or UV luminosity. However, we should point out that at 
z$\sim$ 1, 3, 4, and 5, the SFR$_X$/SFR$_{UV, uncorr}$ $\sim$ 4.5-5 for LBGs,  consistent with the dust extinction values from other methods (Nandra et al. 2002, 
Lehmer et al. 2005, Reddy \& Steidel 2004). Note that the \lya\ luminosity functions of LAEs do not evolve during 3$\leq$ z $\leq$ 6 (e.g., see Ouchi et al. 2008), while
the relative numbers of LAEs to LBGs increased with redshift from z$\sim$ 3 to 6.5 (e.g., Clement et al. 2011), the evolution trend of global f$_{esc}^{Lya}$
could be explained as the evolution of relative numbers of LAEs to LBGs at different redshift, with little implication on the \lya\ escape mechanism  for LAEs only. 
X-ray constraints on the \lyaesc\ of LAEs are independent of dust-extinction law, and we have found the same lower limits of \lyaesc\ for LAEs at z$\sim$ 0.3 and 1, 
and LAEs at z = 2.1 and 3.2 with different average SFR$_{Ly\alpha}$ values. However, a larger sample of LAEs with deep X-ray observation is needed to give a 
constraint on the evolution of \lyaesc\ in LAE only. 


\subsection{Implications from simulated SFR$_X$-L$_X$ relationship}

We should note that the X-ray radiation from galaxies is  predicted to be relatively low for the
youngest stellar populations. 
Mas-Hesse et al.  (2008) predict the soft X-ray to far infrared luminosities ratio in star-forming galaxies from synthesis models. 
They find that the ratio is dependent on the age of the star formation episode for ages $<$ Myr. 
After 30 Myr, the correlation becomes stable and is consistent with SFR$_X$ -- L$_{soft}$ relation of
Ranalli et al. 2003.  
In the hard X-ray band, Mas-Hesse \& Cervino (1999) predicted that a few HMXB should be active in starbursts 
that are older than 5-6 Myr, contributing a few times 10$^{38}$ erg s$^{-1}$ to the total X-ray luminosity.
Recent {\it Chandra} studies in the interacting galaxy pair NGC 4038/4039 (the Antennae; Rangelov et al. in prep) found that 22 of 82 X-ray binaries are coincident or nearly coincident 
with star clusters.  The ages of these clusters were estimated by comparing their multi-band colors with predictions from stellar evolutionary models. 
They found 14 of the 22 coincident sources are hosted by star clusters with ages of $\sim$6 Myr or less. 
So the HMXBs in star-forming galaxies might form earlier than suggested in the 
Mas-Hesse \& Cervino (1999) model.
The estimates of LAE ages are based on SED fitting and are quite uncertain. Such results
have been reported at z = 2.1 (Guaita et al. 2011), 2.25 (Nilsson et al. 2009), 3.1 (Gawiser et al. 2007), 
4.5 (Finkelstein et al. 2008) and z=4.0-5.5 (Pirzkal et al. 2007).  All the results show that the best fit age parameter is 20$\sim$40 Myr, 
but extends to $\sim$0.1-1 Gyr at the 68\% confidence level.
Given this range of best fit ages, we expect that a minority of the LAEs in our sample will be
younger than 15 Myr.  This would imply a modest upward correction to the SFR inferred
from X-rays, and a corresponding correction downwards in the \lyaesc; both corrections might reasonably be factors of $\sim 1.2$--$1.5$.


\section{Conclusions:} From the 4 Ms X-ray Chandra image of CDF-S, we find that the \lyaesc\ from X-rays for LAEs at z$\sim$ 0.3 are about $\sim$2-7 times larger than that from 
dust corrected UV or H$\alpha$.  We coadded $69$ \lya\ emitting galaxies between redshifts $2< z <6.5$. 
None of these galaxies were individually detected. The absence of signal in the coadded image implies an average flux of less than 
7.6$\times$10$^{-19}$ ergs cm$^{-2}$ s$^{-1}$ (1-sigma). This implies that the SFRs in these galaxies are quite modest, 
as indicated by the \lya\ line emission. And the ratio of the average \lya\ line intensity to the upper limits of X-ray flux  constrain
the \lya\ escape fraction of \lyaesc\ $>$ 17\% at 84\% confidence level.  
\\
\\

{\it Acknowledgements: }
We would like to thank the anonymous referee whose comments improved this manuscript significantly. 
ZYZ would like to thank the support from the China Scholarship Council (CSC No. 2009634062) and Arizona State University.  
SM \& JER are supported by the United States National Science Foundation grant AST-0808165. 
The work of JXW is supported by National Basic Research Program of China (973 program, Grant No. 2007CB815404), and Chinese National Science Foundation (Grant No. 10825312). 
\clearpage

\begin{deluxetable}{cccccccccc}
\rotate
\tablecaption{GALEX and CHANDRA observation of three $z$ $\sim$ 0.3 LAEs in the central CDF-S region.}
\tablecolumns{9} \tablewidth{0pt} 
\startdata \hline\hline
  Name & R.A. & Dec. & redshift &   $log$L$_{Ly\alpha}$ &   $^a$EW(Ly$\alpha$) &  SFR$_{Lya}$ &$log$L$_{0.5-2}$  &  $log$L$_{2-10}$ (3$\sigma$)  & $^b$SFR$_X$ \\\hline
  (1)        & (2)     & (3)        & (4)      &      (5)                            &     (6)                    &       (7)               &     (8)            & (9)     & (10) \\ \hline
  GALEX0332-2744 & 53.08000 & -27.74589 & 0.217 & 41.38 & 17 &    0.22      &   39.6 &   $<$40.4     &   1.0       \\
  GALEX0333-2753 & 53.23600 & -27.88797 &  0.365 & 42.32 & 23 &        2.0  &   41.0 & $<$41.3        &   21.7       \\
  GALEX0333-2744 &  53.28029 &  -27.74242 &  0.220  &  41.55 & 12 &       0.3    &   40.7 & $<$41.1      &      10.9    \\ 
\enddata
\tablenotetext{a}{ The values of EW and L$_{Ly\alpha}$ are from Deharveng et al. 2008 and Cowie et al. 2011.  }
\tablenotetext{b}{ The SFR$_{X}$ are calculated with  L$_{soft}$-SFR$_X$ relation of Ranalli et al. 2003.} 
\label{z03lae}
 \end{deluxetable}

\begin{deluxetable}{ccc|cccccc|ccc|cc}
\rotate
\tablecaption{Stacking results of undetected LAEs located in the CDF-S and ECDF-S Field with off-axis-angle $<$ 6\arcmin.}
\tablecolumns{16} \tablewidth{0pt} 
\startdata 
\hline

Redshift & Number$^{a,d}$   &  Time     &      \multicolumn{6}{c}{X-ray COUNTS$^{b}$} & \multicolumn{3}{|c|}{F$_X$($>1\sigma$)$^{c} $} & {lg$_{10}$(L$_X$($>1\sigma$))$^{c}$} & SFR$_X(>1\sigma)^{c}$  \\
        &         & Ms   & Tot$_{S}$ & Net$_{S}$  & Tot$_{H}$ &  Net$_{H}$ &  Tot$_{05-7}$ & Net$_{05-7}$ &  F$_{soft}$ &F$_{hard}$ & F$_{full}$  & L$_{2-10 keV}$  &  (M$_\sun$ yr$^{-1}$)  \\\hline
          \multicolumn{14}{c}{CDF-S only, 50 \% PSF extraction} \\ \hline
   0.3 & 3     &  9.8  &  7.     & 1.3      &  13. & -0.8  &   20. &  0.5    &  0.69 & 2.28 & 1.33  & 39.8  & 1.3 \\
 2.1  &  12  &  39.1  &  20.  &  -4.3  &  71.  &  11.5  &  91.  &  7.1  &  0.19  &  2.54  &  0.98  &  40.8 & 14   \\
3.1  &  18  &  59.2  &  50.  &  -4.9  &  120.  &  -8.2  &  170.  &  -13.0  &  0.18  &  0.96  &  0.51  &  41.2  & 32 \\
3.2  &  23  &  73.9  &  54.  &  -9.4  &  145.  &  -3.5  &  199.  &  -12.9  &  0.15  &  0.84  &  0.44  &  41.1  & 28 \\
4.5  &  12  &  39.3  &  30.  &  2.5  &  59.  &  -6.8  &  89.  &  -4.3  &  0.31  &  1.05  &  0.58  &  41.8  & 139 \\
5.7  &  3  &  9.8  &  12.  &  -0.6  &  34.  &  5.9  &  46.  &  5.4  &  0.62  &  6.19  &  2.90  &  42.3  & 439 \\
6.5  &  1  &  3.3  &  3.  &  -3.6  &  16.  &  0.6  &  19.  &  -2.9  &  1.17  &  8.16  &  3.53  &  42.7  & 1002  \\     \hline  
           \multicolumn{13}{c}{CDF-S + ECDF-S, 50 \% PSF extraction} \\ \hline      
 0.3  &  12  &  11.6 & 19  &  6.4  &  31  &  -0.1  &  50  &  6.3  &  1.3  &  2.7  &  2.7 &  39.9 & 1.6  \\
 1.0 & 2    &  0.4  & 1  & -1.2 & 9 & 3.4 & 10 & 2.2 &  7.3 & 83.9 & 32.9 & 41.6 & 84 \\
2.1  &  122  &  72.2  &  49.  &  1.9  &  125.  &  11.7  &  169.  &  9.1  &  0.18  &  1.54  &  0.68  &  40.8  & 14 \\
3.1  &  118  &  85.8  &  71.  &  -6.2  &  169.  &  -3.4  &  234.  &  -12.4  &  0.14  &  0.76  &  0.40  &  41.1 & 28  \\
3.2  &  24  &  74.9  &  55.  &  -9.1  &  159.  &  5.9  &  216.  &  -2.7  &  0.15  &  1.24  &  0.45  &  41.1  & 28 \\
4.5  &  64  &  49.8  &  44.  &  6.8  &  93.  &  6.8  &  136.  &  13.7  &  0.38  &  1.65  &  1.13  &  41.9  & 160 \\
5.7  &  17  &  12.6  &  14.  &  -0.0  &  37.  &  5.6  &  51.  &  5.6  &  0.50  &  4.73  &  2.33  &  42.3 & 440  \\
6.5  &  6  &  4.5  &  4.  &  -3.0  &  19.  &  2.8  &  23.  &  -0.2  &  0.93  &  8.55  &  2.79  &  42.6  & 876 \\         \hline    
\enddata
\tablenotetext{a}{ Number of LAEs located in central CDF-S and ECDF-S region selected for stacking analysis. Reference: LAEs at z $\sim$ 0.3 from Deharveng et al. (2008) and Cowie et al. (2011), z $\sim$ 1 from Cowie et al. (2011), z = 2.1 from Guaita et al. (2010), z = 3.1 from Gronwall et al. (2007) and Ciardvllo et al. (2011), z = 3.15 from Nilsson et al. (2007), z = 4.5 from Finkelstein et al. (2009), z = 5.7 from Wang et al (2005)  and z = 6.5 from Rhoads et al. in prep. } 
\tablenotetext{b}{Notice that the number of counts are extracted from their 50\% PSFs and summed up, 
and not corrected for the apertures.} 
\tablenotetext{c}{ The $1\sigma$ flux limits are obtained by first calculating the 
$1 \sigma$ upper limit on counts as $Net$+1$\times(\sqrt{Tot+0.75}+1)/(\hbox{PSF-fraction})$.  Here ``Net'' and ``Tot'' are the net and total counts in the 50\% PSF region, 
respectively, while ``PSF-fraction'' here is 50\%.  The counts are then divided by 
effective integration time and multiplied by the count-rate to flux convertsion factor. 
The tabulated flux limits are in units of 10$^{-17}$ \fluxunit. The SFR$_X$ for all LAEs at z$>$2 are converted with the relationship of Ranalli et al. 2003 from 1-$\sigma$ upper limit on L$_{2-10 keV}$. } 
\tablenotetext{d}{Due to the rare number of GALEX selected LAEs, we enlarge the selection area to off-axis-angle $\leq$ 9" for z$\sim$ 0.3 and 1.    Only 2 z$\sim$1 LAEs were located in ECDF-S region, and did not show any X-ray detection. }
\label{stkdata}
 \end{deluxetable}

\begin{deluxetable}{ccc|ccc|c|cc}
\rotate
\tablecaption{The LAE samples used in this work and their average SFRs and average \lya\ escape fractions in the 4Ms CDF-S region.}
\tablecolumns{6} \tablewidth{0pt} 
\startdata \hline
Redshift    & Lg L$(Ly\alpha)_{limit}$$^b$ & EW($Ly\alpha$)$_{rest, limit}$$^b$& SFR$^c_X$ & SFR$^c_{Ly\alpha}$ &  SFR$^c_{UV}$ & A$^d_{1200}$  & f$^{esc}_{(Ly\alpha / UV\_corr)^e}$ & f$^{esc}_{(Ly\alpha / X)^e}$ \\ 
 & & \AA\ & \multicolumn{3}{|c|}{ (M$_\sun$ yr$^{-1}$) } & (mag)  & (\%) & (\%)  \\ \hline\hline
0.217$^a$ & -- & -- &1.0  & 0.22 & 2.9& 1.25$\pm$0.35 & 3.3$^{+1.2}_{-1.0}$ & 22.0  \\
0.220$^a$ & -- & -- &10.9 & 0.3 & .. & ..       &  ..       & 2.8 \\
0.374$^a$   & -- & -- & 21.7 & 2 & 6.2 & 2.00$\pm$0.00 & 5.1 & 9.2  \\ \hline
$\sim$0.3   & 41.2 & 20 &  $<$1.6  & 0.9 & .. & .. & .. & $>$56.3 \\
$\sim$1.0 & 42.6  & 20 & $<$ 84  & 5.7 & .. & .. & .. & $>$ 6.8 \\ \hline\hline
2.1   & 41.8 & 20 & $<$14 & 1.9 & 2.4 & 1.2$^{+0.5}_{-1.2}$ & 26$^{+53}_{-9}$& $>$14 \\
3.1  & 42.1 & 20 & $<$28 & 2.6 & 5.3 &  $<$ 0.6 & 28$^{+21}_{-0}$ &  $>$9.4\\
3.2  & 42.9 & 22 & $<$ 28 &  2.3 & 4.2 & 0.8$^{+0.3}_{-0.5}$ & 26$^{+16}_{-6}$ & $>$8.3\\
4.5  & 42.6 & 16 & $<$ 139 & 6.0 & 24 & 1.5$^{+3}_{-1.1}$  & 6$^{+11}_{-5.6}$& $>$4.3\\
5.7   & 42.8 & 11 &  $<$ 439 & 5.5 & ... & ... & ... & $>$1.3  \\
6.5  &... & ... &  $<$876 & 5 & ... & ...  & ...  & $>$0.57\\
\enddata
\tablenotetext{a}{ The three LAEs at z $\sim$ 0.3 are individually detected, and the other results of LAEs  are stacked or averaged.}
\tablenotetext{b}{ The selection limits on L($Ly\alpha$) and EW($Ly\alpha$) for our sample. Note that the LAEs at z$\sim$ 0.3 and 1 are selected through broadband drop-out first, then applied spectroscopic observation. Their EW($Ly\alpha$) are estimated from their optical spectra, and select EW$_{rest}>$ 20 \AA\ for comparison with LAEs at higher redshifts. Other surveys for LAEs at z$>$ 2 are selected through narrowband selection, with nearly same selection criteria on EW($Ly\alpha$). Although LAEs at z = 4.5 show threshold of \lya\ EW $>$ 16 \AA,  only $<$ 6\% of them have EW$<$ 20 \AA. }
\tablenotetext{c}{ The SFR$_X$ for all LAEs at z$>$2 are converted with the relationship of Ranalli et al. 2003 from 1-$\sigma$ upper limit on L$_{2-10 keV}$. The SFR$_{Ly\alpha}$ and SFR$_{UV}$ are converted with Kennicutt 1998.} 
\tablenotetext{d}{ The A$_{1200}$ were converted from the dust properties of different SED fitting papers under Calzetti et al. (2000) dust law: Finkelstein et al. 2010 for z$\sim$0.3 LAEs, Guaita et al. 2011 for z$\sim$2.1 LAEs, Gawiser et al. 2007 for z$\sim$3.1 LAEs, Nilsson et al. 2007 for z$\sim$3.15 LAEs, and Finkelstein et al. 2008 for z$\sim$4.5 LAEs.}
\tablenotetext{e}{ See text (chapter \S 4.2 and \S 4.4 for the definitions of the \lya\ escape fraction estimated from extinction-corrected UV flux and X-ray luminosity. }
\label{xsfr}
 \end{deluxetable}
 

  \clearpage

\end{document}